\documentclass[twocolumn]{aastex631}

\begin{document}

\title{Probing the Nature of the First Galaxies with JWST and ALMA}

\correspondingauthor{Mar\'{\i}a Emilia De Rossi}
\email{mariaemilia.dr@gmail.com}

\author[0000-0002-4575-6886]{Mar\'{\i}a Emilia De Rossi}
\affiliation{Universidad de Buenos Aires, Facultad de Ciencias Exactas y Naturales y Ciclo B\'asico Com\'un,
Buenos Aires, Argentina}
\affiliation{CONICET-Universidad de Buenos Aires, Instituto de Astronom\'{\i}a y F\'{\i}sica del Espacio (IAFE),
Buenos Aires, Argentina}

\author[0000-0003-0212-2979]{Volker Bromm}
\affiliation{Department of Astronomy, 
University of Texas at Austin, 2515 Speedway, Stop C1400, Austin, TX 78712, USA}
\affiliation{Weinberg Institute for Theoretical Physics, 
University of Texas at Austin, Austin, TX 78712, USA}







\begin{abstract}
By implementing a model of primordial dust emission, we predict {\em dust-continuum} fluxes for massive galaxy sources similar to those recently detected by JWST at $z \ga 7$.
Current upper flux limits, obtained with ALMA for some of these sources, can constrain gas metallicity and dust fraction of the first galaxies.
Encouragingly, if assuming expected properties for {\em typical} first galaxies (i.e., dust-to-metal mass ratio: $D/M = 5 \times 10^{-3}$, gas metallicity: $Z_{\rm g} = 5 \times 10^{-3}~Z_\sun$, star formation efficiency: $\eta = 0.01$), model far-infrared (FIR) fluxes are consistent with current upper flux limits inferred from ALMA bands 6 and 7 ($\la 10^4$ nJy).  
Such low $D/M$ values and metallicities are in agreement with some scenarios proposed in the literature to explain the non-detection of the FIR dust continuum for high-$z$ JWST galaxy candidates.  On the other hand, higher values of model parameters $D/M$ ($\ga 0.06$) and $Z_{\rm g}$ ($\ga 5 \times 10^{-2}~Z_\sun$) are ruled out by observational data, unless a higher $\eta$ is assumed.   
According to our findings, ALMA multi-band observations could constrain the dust chemistry and dust grain size distribution in the early universe. In this context, future observational challenges would involve not only reaching higher FIR sensitivities, but also increasing the wavelength coverage by exploring distinct ALMA bands.  
\end{abstract}



\section{Introduction} \label{sec:introduction}
With the successful launch of the JWST, astronomy has entered an exciting period of extending the frontiers of what we know about the universe. Initial results from JWST imaging hint at a surprising abundance of massive galaxies already briefly after cosmic dawn, at redshifts $z\gtrsim 10$ \citep[e.g.,][]{Finkelstein2022,Harikane2022,Labbe2022,Naidu2022, Adams2023, Atek2022}. If confirmed by spectroscopic follow-up, such early emergence of massive galaxies may seriously challenge the $\Lambda$CDM standard model of cosmological structure formation \citep{Boylan2022}, with no obvious way to accelerate early galaxy formation \citep[e.g.,][]{Klypin2021,LiuBromm2022}. 

To firm up the redshift estimates \citep{Fujimoto2022}, and to elucidate the physical nature of the sources \citep{BrommYoshida2011,Kohandel2023}, longer wavelength observations with the Atacama Large Millimeter/sub-mm Array (ALMA) in the far infrared (FIR) and sub-mm bands are vital. One key target for ALMA follow-up are the strong fine-structure FIR cooling lines of ionized oxygen and carbon, [OIII] and [CII], so far resulting only in upper limits on the line fluxes \citep[e.g.,][]{Bakx2022,Yoon2022,Kaasinen2022}. Similarly, attempts to directly probe the dust continuum in the JWST galaxies have led to only non-detections until now \citep{Fujimoto2022}.  

The dust content within the first galaxies, regarding overall mass fractions and detailed chemical make-up, is important in shaping their spectral energy distributions (SEDs) and morphologies \citep[e.g.,][]{Jaacks2018,Ferrara2022}, expressed in quantities such as the UV slope, $\beta_{\rm UV}$, or the interstellar medium (ISM) dust extinction\footnote{The UV slope is defined through a power-law fit to the respective portion of the SED, such that $f_{\lambda}\propto \lambda^{\beta_{\rm UV}}$.}. Dust extinction is also a key effect in determining the escape fraction of ionizing radiation as well as of resonantly scattered Lyman-$\alpha$ photons \citep[e.g.,][]{Smith2019}. Furthermore, the nature of dust in the first galaxies is indicative of dust production channels in the early universe, where timescales are favoring more rapid pathways, such as supernova (SN) explosions \citep[e.g.,][]{ji2014}. Furthermore, the compositional nature of the dust in the first galaxies can constrain the SN enrichment from the first stars, and thus indirectly also probe their initial mass function \citep[e.g.,][]{Gall2011}.

Given the importance of the dust content in the first galaxies, we here specifically explore the dust emission for JWST sources at $z \ga 7$. Assuming that JWST has determined the galaxy's stellar mass, we predict the corresponding dust fluxes, thus assessing the ALMA observability. We also check for consistency with current upper limits, thus testing the overall theoretical framework for early galaxy formation in a way that is complementary to other such tests.

\section{Methodology} \label{sec:methodology}
We apply the dust model developed by \citet{derossi2017} to estimate the {\em dust continuum FIR signatures} associated with primeval massive galaxies, such as those recently detected by the JWST at $z \gtrsim 7$. This model has been successful at predicting the FIR fluxes of first massive galaxies at $z \gtrsim 5$ \citep{derossi2018}. For the convenience of the reader, we briefly summarize our methodology below; a more detailed description can be found in \citet{derossi2017, derossi2019}. 

\subsection{Galaxy Formation Model}
\label{sec:galaxy_formation_model}

In our model, a first galaxy consists of a central Population~II (Pop~II) compact stellar cluster, inhabiting a virialized dark-matter halo in a $\Lambda\mathrm{CDM}$ universe.\footnote{With parameters $h$ = 0.67,
${\Omega}_{\rm b}$ = 0.049,
${\Omega}_{\rm m}$ = 0.32,
${\Omega}_{\Lambda}$ = 0.68
\citep{planck2014}.} The stellar source is surrounded by an ISM, where gas and dust are mixed. For simplicity, we assume spherical symmetry and do not consider any extended distribution of halo stars.
We adopt a \citet{burkert1995} gas density profile, imposing a total-to-baryon ratio of the order of the cosmic mean ($\Omega_{\rm b} / \Omega_{\rm m}$).\footnote{As discussed in  \citet{derossi2017}, similar trends are obtained if using other profiles, such as an isothermal or Navarro–Frenk–White form.}
For estimating the stellar mass ($M_*$), we consider a conservative star formation efficiency of $\eta = M_* / (M_{\rm g} + M_*) = 0.01$, where $M_{\rm g}$ is the gas mass \citep[e.g.][]{greif2006, mitchell2015}, but we also explore higher $\eta$ values, as shown below.
In order to generate the SED for the stellar component, we use Yggdrasil model grids \citep{zackrisson2011}, implementing the lowest available stellar metallicity, $Z_* \approx 3 \times 10^{-2} Z_{\sun}$, and a stellar age $\tau = 0.01$ Myr.

\subsection{Dust Physics in the First Galaxies}
\label{sec:dust_physics}
We compute the dust spatial distribution within each halo from the gas-density profile, assuming a
dust-to-metal mass ratio $D/M = M_{\rm d} / M_{Z} = 5 \times 10^{-3}$, and a gas metallicity of 
$Z_{\rm g} = 5 \times 10^{-3} Z_{\sun}$ \citep{derossi2017, derossi2019}. This parameter choice predicts a total dust mass in agreement with the scaling in I Zw 18, which is a
local analog of extremely metal-poor galaxies at high $z$.  In addition, we test variations of model parameters to evaluate the sensitivity of our predictions to them.

Following \citet{ji2014}, we describe dust chemical composition by applying the silicon-based models of \citet{cherchneff2010}: UM-ND-20, UM-ND-
170, UM-D-20, UM-D-170, M-ND-20, M-ND-170, M-D-20, and
M-D-170.\footnote{Dust model notation follows that in \citet{cherchneff2010}, i.e., UM: unmixed, M: mixed; ND: non-depleted, D: depleted;
170: 170 ${\rm M}_{\sun}$ progenitor, 20: 20 ${\rm M}_{\sun}$ progenitor.}
We note that there is a debate regarding the role of carbon-based dust in primeval galaxies.  As discussed in \citet{derossi2017, derossi2019}, a moderate contribution of carbon dust in our models would only drive slight changes in dust temperature ($T_{\rm d}$) and an enhancement of dust emission by a factor of a few, but the main predicted trends would be preserved.
With respect to the grain size distribution, we consider the so-called ``standard'' and ``shock'' prescriptions used by \citet{ji2014}. The former is similar to the Milky Way one \citep{pollack1994}, while the latter, which predicts smaller dust grains, is based on \citet{bianchi2007} and approximates the effects of running a post-SN reverse shock through newly created dust.
We do not try to assess which dust model is more realistic, but instead use all of them to evaluate the impact of dust chemistry and grain sizes on our findings.

For estimating $T_{\rm d}$, we assume thermal equilibrium between dust cooling and heating rates, with the latter driven mainly by stellar photo-heating and secondly by dust-gas collisions. The cosmic microwave background sets a temperature floor as it is not thermodynamically possible to radiatively cool below it.

\begin{figure*}
        \begin{center}
\plottwo{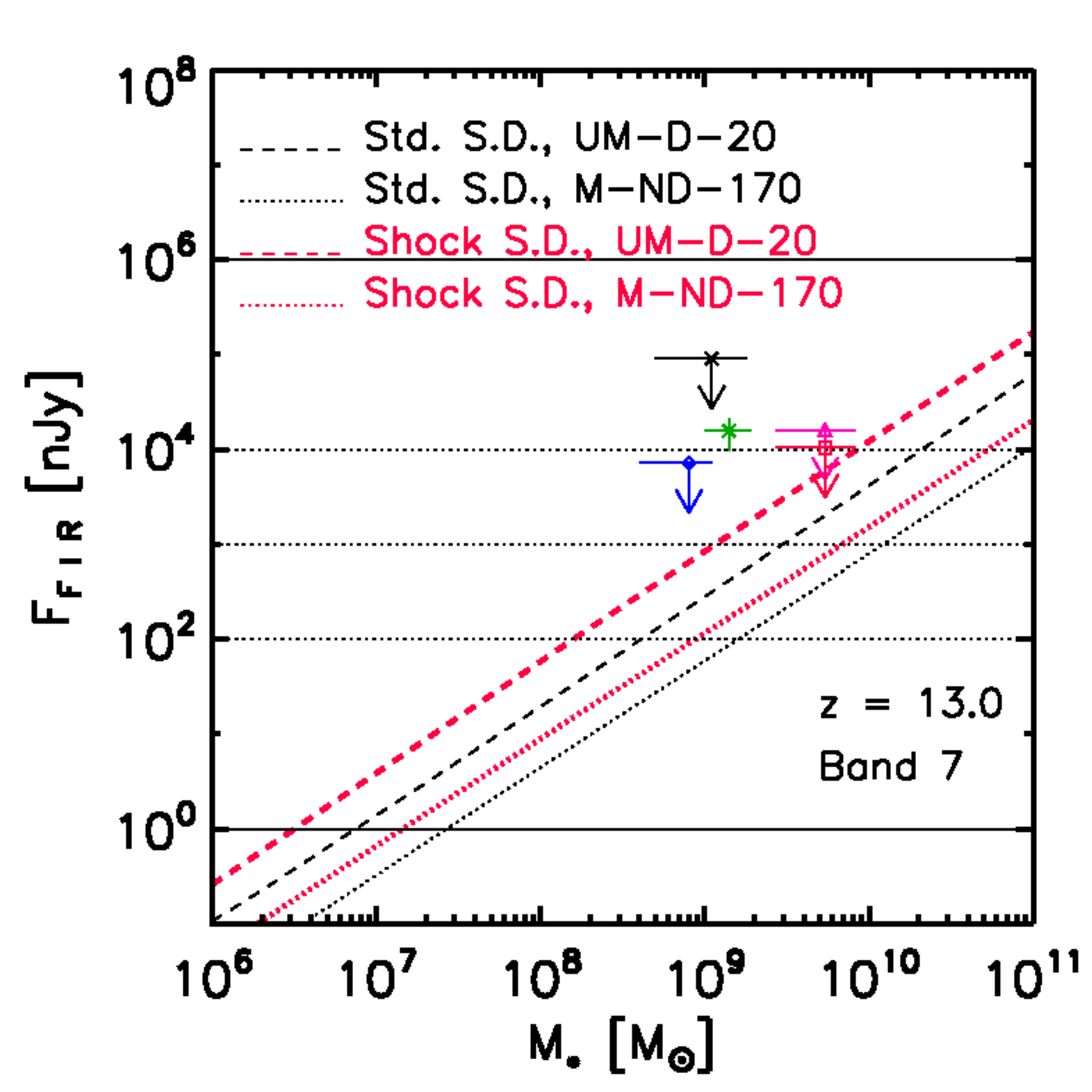}{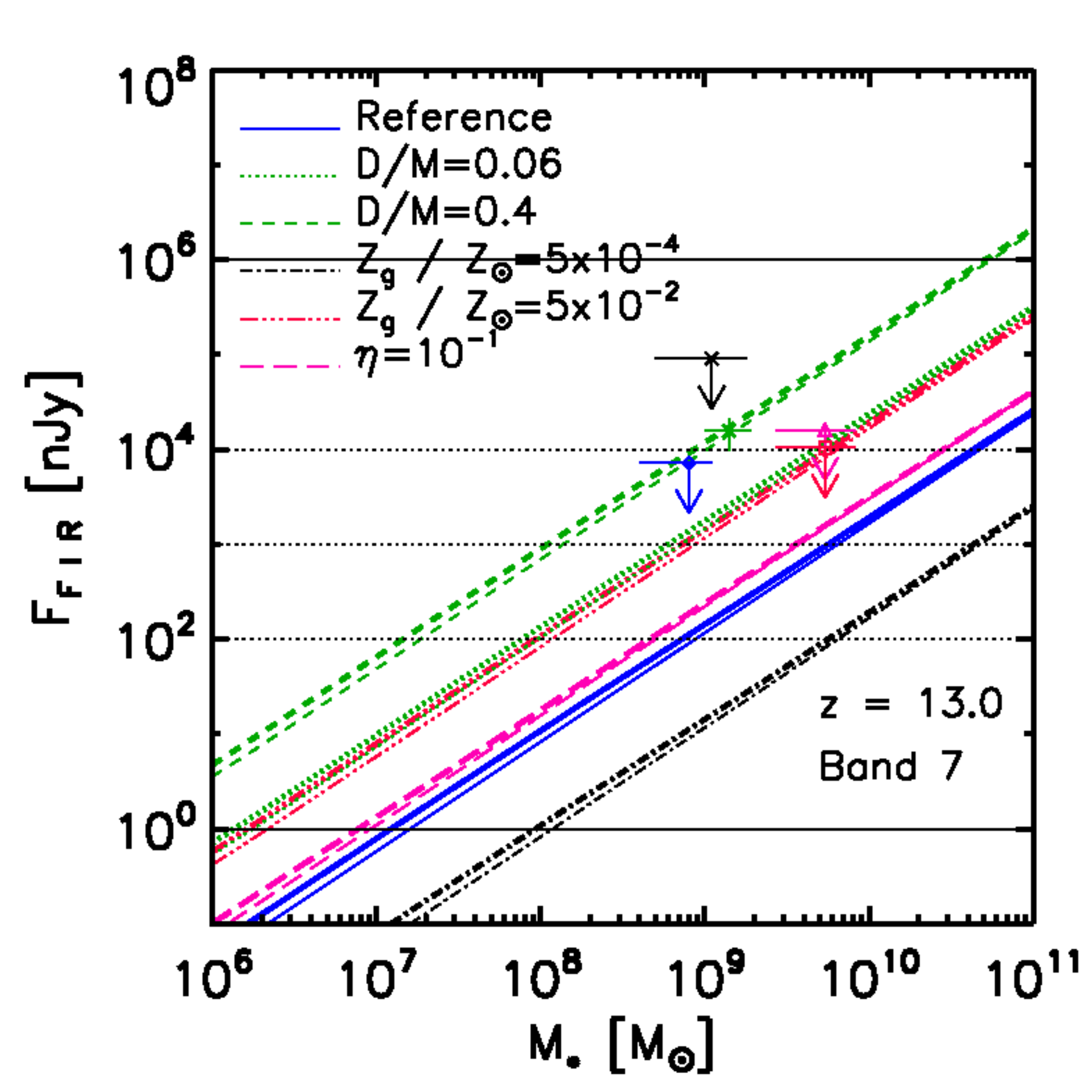}
        \end{center}
    \caption{
	    Average FIR flux associated with ALMA band 7, as a function of $M_{*}$ for model galaxies at $z=13$. {\it Left-hand panel}: Comparison between dust models that adopt different grain size distributions (standard and shock, shown in black and red, respectively) and chemical compositions (UM-D-20 and M-ND-170, shown with dashed and dotted lines, respectively; for a given size distribution, all other chemical models lead to an intermediate behavior). {\it Right-hand panel}: Effects of varying the dust-to-metal ratio ($D/M$), gas-phase metallicity ($Z_{\rm g}$), and star formation efficiency ($\eta$) of dark matter halos; results correspond to our default dust chemistry (UM-ND-20), assuming a standard (thin lines) and shock (thick lines) grain size distribution.
        Different symbols depict upper limits (arrows) and a tentative value (green symbol) discussed in the literature (see the text for details).  Within our model uncertainties, current upper limits are not consistent with our maximum $D/M$.
    }
    \label{fig:fig1}
\end{figure*}

\subsection{JWST/ALMA signatures}
\label{sec:jwst_alma_signatures}
As mentioned above, we aim at predicting ALMA fluxes for galaxies recently discovered by the JWST/NIR instruments at $z \gtrsim 7$.  
Since our model is specially designed to study the FIR radiation from first galaxies and does not implement all the required processes (such as nebular emission and the reprocessing of $\mathrm{Ly}\alpha$ photons in the intergalactic medium) for modelling observed fluxes in the NIR, we assume that $M_*$ has been independently derived from the JWST/NIR measurements.

In order to obtain ALMA fluxes, we first calculate the dust emissivity per unit mass ($j_{\nu}$) by
applying Kirchhoff's law for the $T_{\rm d}$ profile derived from our model.

Then, the total specific {\em dust} luminosity $L_{\nu ,{\rm em}}$ corresponding to a given galaxy is estimated by integrating $j_{\nu}$ out to the virial radius ($R_{\rm vir}$) of its host halo

The observed {\em dust} specific flux $f_{\nu , {\rm obs}}$ from the model source is calculated as:

\begin{equation}
f_{\nu , {\rm obs}} = (1 + z)  \frac{L_{\nu ,{\rm em}}}{4 \pi {d_{L}}^2}
\mbox{\, ,}
\end{equation}
where $d_{L}$ is the luminosity distance to a galaxy at redshift $z$.
Finally, we compute the average specific flux observed over a given ALMA band via:
\begin{equation}
\label{eq:average_flux}
F_{\rm FIR} = \frac{\int_{{\nu}_{i}}^{{\nu}_{f}} f_{\nu , {\rm obs}} \ {\rm d}{{\nu}_{\rm obs}}}{{{\nu}_{f}}-{{\nu}_{i}}},
\end{equation}
where ${{\nu}_{i}}$ and ${{\nu}_{f}}$ delineate the frequency range associated to
that band. We consider all currently available bands (Band 3: 84-116 GHz, Band 4: 125-163 GHz, 
Band 5: 163-211 GHz, Band 6: 211-275 GHz, Band 7: 275-373 GHz, Band 8: 385-500 GHz, 
Band 9: 602-720 GHz, Band 10: 787-950 GHz).

\begin{figure*}
        \begin{center}
\plottwo{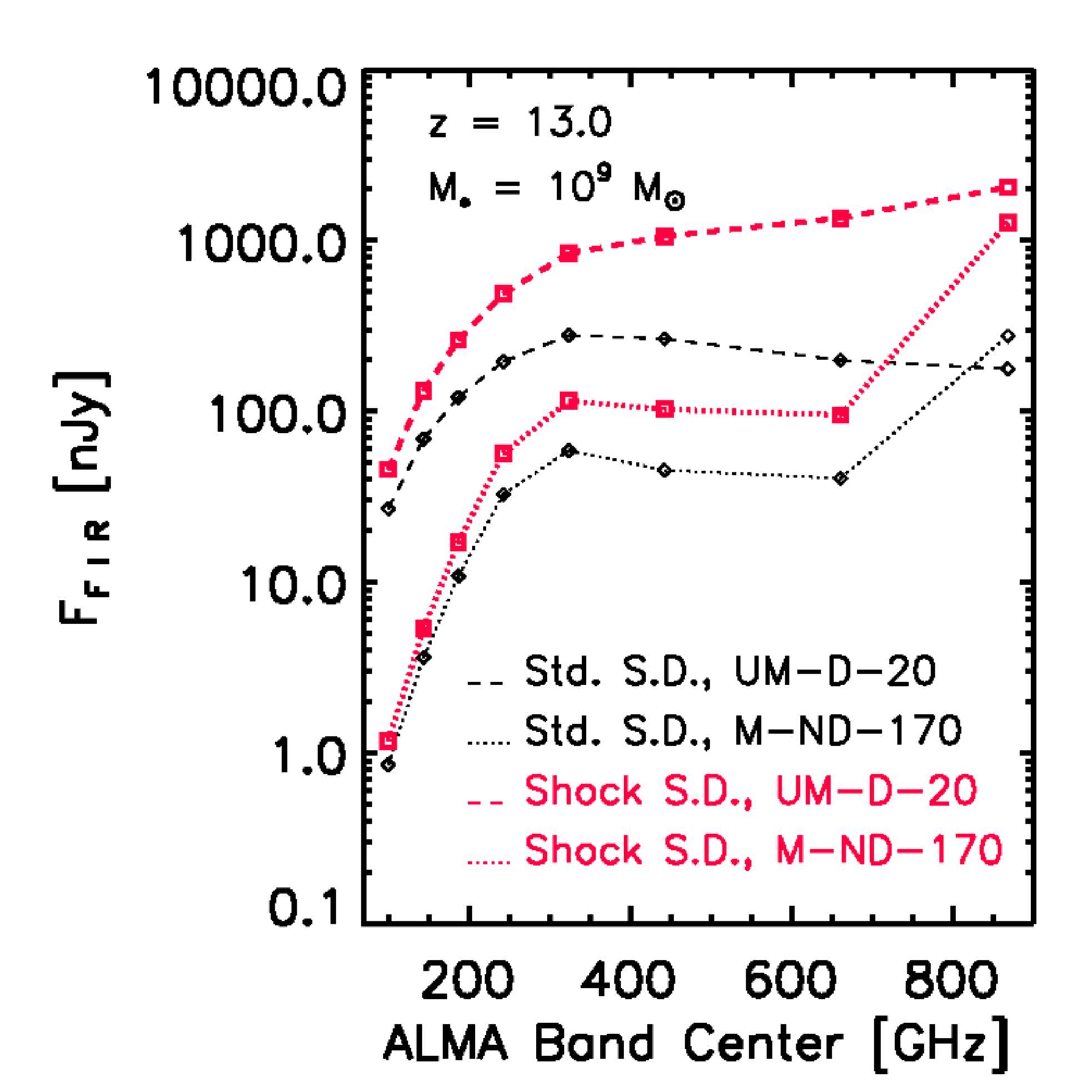}{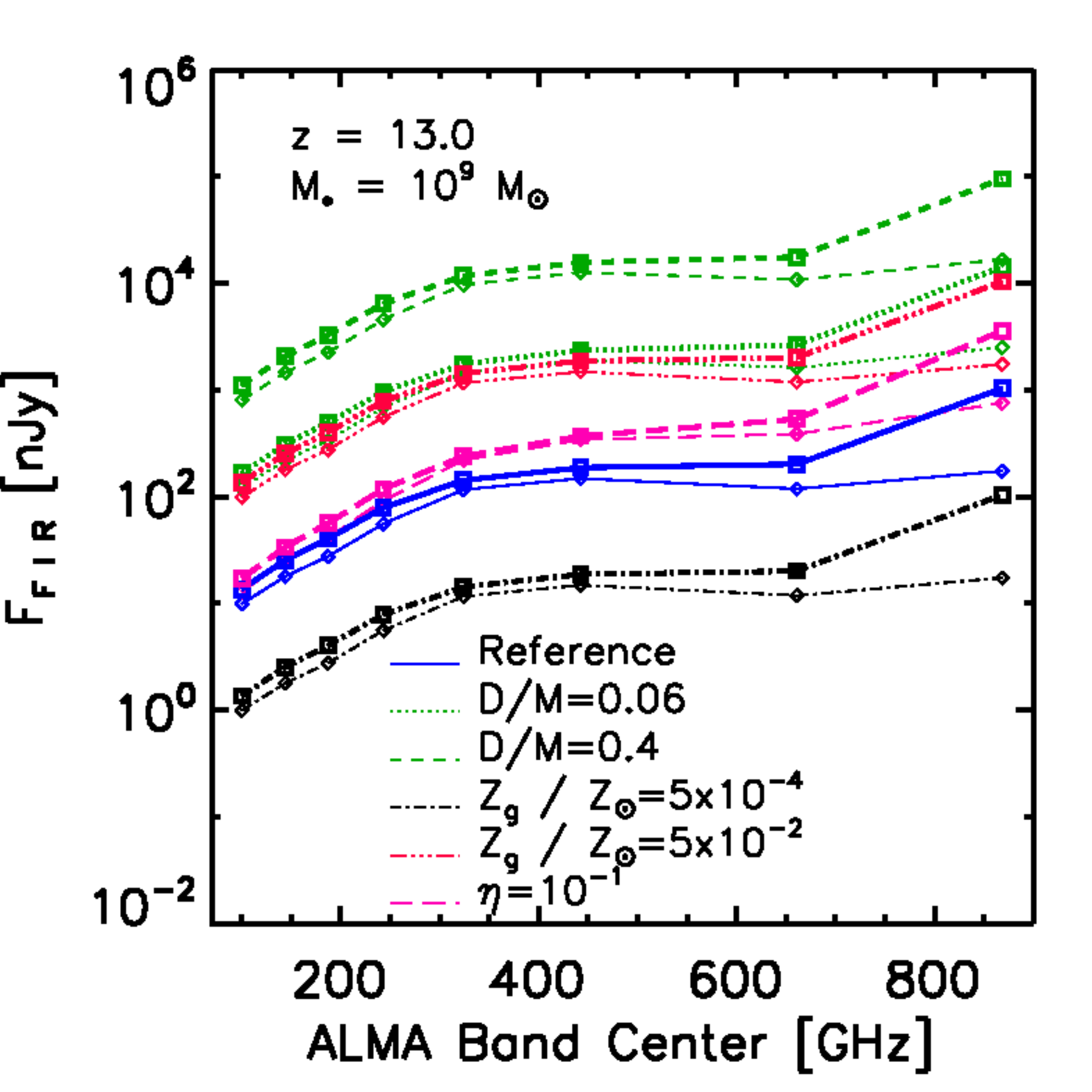}
        \end{center}
    \caption{
	    Average FIR flux within available ALMA bands (3\--10) vs. central band frequency, for a model source of $M_* = 10^{9}~{\rm M}_{\odot}$ at $z=13$. Results correspond to the same models shown in Fig.~\ref{fig:fig1}. {\it Left-hand panel}: Comparison between dust models that adopt different grain size distributions (standard vs. shock) and chemical compositions (UM-D-20 vs. M-ND-170), for a given size distribution; all other chemical models generate an intermediate behavior. {\it Right-hand panel}: Effects of varying the dust-to-metal ratio ($D/M$), gas-phase metallicity ($Z_{\rm g}$), and star formation efficiency ($\eta$). Here, results correspond to our default dust chemistry (UM-ND-20), again assuming a standard (thin lines) and shock (thick lines) grain size distribution.
    }
    \label{fig:fig2}
\end{figure*}

\section{Results}
\label{sec:results}
In Figure~\ref{fig:fig1}, we analyze the average ALMA band 7 flux predicted by our dust model as a function of the stellar mass, $M_*$, of a galaxy located at $z = 13$. 
We show results for different dust models and compare them with data reported in \citet[][see their tables 1 and 3]{Fujimoto2022}, corresponding to $z \sim 11 - 17$ galaxy candidates observed with ALMA bands 6 and 7.
Different symbols depict select observed sources: S5-z17-1 (black cross, band 7, \citealt{Fujimoto2022}), GHZ1/GLz11 (green asterisk, band 7, \citealt{Yoon2022}), GHZ2/GLz13 (blue circle, band 6, \citealt{Bakx2022}, \citealt{Popping2023}), and HD1 (pink triangle, band 6, \citealt{Harikane2022b}; red square, band 4, \citealt{Kaasinen2022}), plausibly located at $z\approx 18.41$, $10.87$, $12.43$ and $15.39$, respectively \citep{Fujimoto2022}. 
With the only exception of GHZ1/GLz11, for which a tentative value is represented, all other sources are marked with arrows as only upper flux limits are available for the associated ALMA bands.
We notice that, if using band 6 for estimating our model fluxes, we predict only a slight decrease by $\la 0.3$ dex with respect to fluxes obtained for band 7 (see below). Only the model adopting $\eta = 0.1$ and standard grain size distribution predicts a higher flux decrease of $\approx 0.4$ dex when using band 6.

Model fluxes inferred from different dust chemical compositions and grain size distributions are evaluated in the left panel of Figure~\ref{fig:fig1}. Only chemical patterns with extreme behaviors are plotted (UM-D-20, M-ND-170); for all other cases, intermediate trends are obtained. We see that, at a given $M_*$ and for the same dust chemistry, higher fluxes are
predicted for the shock size distribution (i.e. smaller dust grains), in agreement with previous findings by \citet{derossi2019}.
Encouragingly, the predicted FIR fluxes are consistent with current upper limits estimated from ALMA measurements, regardless of the specific dust properties adopted.  We verified that very similar trends are obtained from our model if varying $z$ along the whole observed redshift range ($\sim 11-17$).  As shown below, our model also predicts lower fluxes for band 4 than for band 6, which is consistent with the decrease of the upper flux limits observationally determined for HD1 between such bands.

We also explore the impact of changing the model dust-to-metal mass ratio, gas metallicity and star formation efficiency with respect to our adopted fiducial values ($D/M = 5 \times 10^{-3}$, $Z_{\rm g} = 5 \times 10^{-3}~Z_\sun$, $\eta = 0.01$, respectively; see Sec.~\ref{sec:methodology}). For this analysis, we employ the default UM-ND-20 dust chemistry implemented in \citet{derossi2019}, which predicts intermediate trends between the UM-D-20 and M-ND-170 extreme cases studied previously. Results are shown in the right panel of Figure~\ref{fig:fig1} for the standard (thin lines) and shock (thick lines) size distributions.  According to our findings, current upper limits for observed FIR fluxes are not consistent with an extremely high $D/M \approx 0.4$. In addition, a less extreme $D/M \approx 0.06$ or a high $Z_{\rm g} \approx 5 \times 10^{\rm -2}~Z_\sun$ are only marginally acceptable, and would be ruled out for certain dust chemical compositions which result in higher fluxes (e.g., UM-D-20, as discussed before). 
We also consider a higher ${\eta} = 0.1$, in which case a $Z_{\rm g} \approx 5 \times 10^{\rm -2}~Z_\sun$ is adopted 
\citep{derossi2019}.\footnote{This assumption corresponds to the closed-box model approximation, which provides an upper limit for the predicted dust fluxes.}
Figure~\ref{fig:fig1} shows that a higher $\eta$ or a lower $Z_{\rm g}$, compared to the values in our reference case, are still in agreement with current upper flux limits.
Similar trends are obtained for other redshifts in the range $z\sim 11 -17$.

In Figure~\ref{fig:fig2}, we approximately mimic an ALMA multi-band analysis for a model galaxy at $z=13$ and with $M_* = 10^9~{M_\sun}$, which are values close to those estimated for the high-$z$ galaxy sources recently detected by JWST. Similarly to Figure~\ref{fig:fig1}, the left panel compares predictions for models that implement different dust chemical compositions and grain size distributions, whereas the right panel evaluates the effects of changing $D/M$, $Z_{\rm g}$ and $\eta$ for the UM-ND-20 chemistry and different grain size distributions.  In general (see left panel), we note that $F_{\rm FIR}$ increases from band 3 to 7, remains almost constant from band 7 to 9, and reaches a higher value at band 10 for some dust models.  However, the exact features of the $F_{\rm FIR}$ vs. waveband relation depend on dust chemistry and grain size distribution, suggesting that its determination through multi-band studies could be crucial to constrain the nature of dust in the early universe. On the other hand, the right panel of Figure~\ref{fig:fig2} shows that the absolute normalization of the $F_{\rm FIR} - {\rm band}$ curve can help to constrain other key galaxy properties such as $D/M$, $Z_{\rm g}$ and $\eta$. 

Finally, we note that vigorous efforts are being made to detect prominent ISM cooling lines with ALMA, such as the [OIII] and [CII] FIR emission lines, trying to obtain robust spectroscopic redshifts for select JWST sources.  
We emphasize that the detection of the dust continuum emission in different ALMA bands would
contribute to better constrain the SEDs of primeval galaxies, which would ideally complement studies searching for the aforementioned emission lines. 

\section{Discussion of Parameter Choices}
We use the ALMA sensitivity calculator\footnote{https://almascience.eso.org/proposing/sensitivity-calculator} to provide a rough estimate of the time required to achieve an instrument sensitivity of the order of our model FIR fluxes.
By selecting the commonly studied ALMA band 7 (band 6), an observing frequency of 346 GHz\footnote{Equivalent to an observed wavelength of $\lambda = 866~\mu {\rm m}$, a good tracer of the dust continuum at early epochs.} (240 GHz), and a default configuration of 43 12 m array antennas,  $\approx 6$ hr ($\approx 3$ hr) and $\approx 26$ days ($\approx 11$ days) are required to reach sensitivities of $\sim 10^4$ and $\sim 10^3$ nJy, respectively.  
In the case of sources shown in Figure~\ref{fig:fig1}, S5-z17-1 was observed for $\approx 16$ min, for example, while an observing time $>10$ hr was employed for GHZ1/GLz11 and GHZ2/GLz13, with the dust continuum not detected for either of them. Our reference model predicts FIR fluxes $\la 10^3$ nJy within the mass range of these sources, suggesting that several days would be needed to reach the sensitivity for direct detection of the dust continuum.  However, very recent observations suggest higher $Z_{\rm g}$ for systems in this mass range at $z>10$, in which case smaller observing times would be required, increasing the probability of being detected with ALMA.

Based on their SED fitting, \citet{curtislake2023} report stellar metallicities $\ga 10^{\rm -2}~Z_\sun$ for a 
few spectroscopically confirmed $z>10$ galaxies with $M_* \sim 4\times 10^7 - 5\times 10^8~{\rm M_\sun}$.  Also from SED fitting, \citet{bunker2023} estimate a nebular gas metallicity $\ga 10^{\rm -1}~Z_\sun$ (which is consistent with the value inferred from an emission line analysis) and $M_* \sim 5\times 10^8$ for the $z>10$ galaxy candidate GN-z11. Our model $F_{\rm FIR}$ fluxes increase with $D/M$ and $Z_{\rm g}$, which are degenerate parameters.  If the sources modeled in Figure~\ref{fig:fig1} were to reach $Z_{\rm g} \ga 10^{\rm -2}~Z_\sun$, the constraints on $F_{\rm FIR}$ would imply $D/M$ ratios significantly lower than our reference value ($5 \times 10^{-3}$). Such extremely low $D/M$ in turn would not be consistent with the observed $Z_{\rm g} - D/M$ relation at $z=0-5$ 
\citep[e.g.,][]{popping2022, peroux2020}. If confirmed, this disagreement may be evidence for significant evolution of the $Z_{\rm g} - D/M$ relation towards higher redshifts, $z\approx 5-10$.
On the other hand, enhancing $\eta$ also drives an increase of
$Z_{\rm g}$ in our model, preserving the agreement with current $F_{\rm FIR}$ observational limits, as well (see Sec.~\ref{sec:results}).

We acknowledge that properties of $z>10$ galaxies are quite uncertain and our reference model adopts very conservative parameters, expected for {\em typical} primeval galaxies, dust-enriched by the very first stars. 
Metallicities $\ga 10^{\rm -2}~Z_\sun$ at $z>10$ might be associated with more evolved systems than the average galaxy population at this epoch, with further observations required for clarification.  
Furthermore, it is worth emphasizing that metallicities at $z>10$, based on SED fitting, involve uncertain assumptions resulting in large error bars. 
More robust metallicity determinations require methods based on spectral lines. However, as discussed in \citet{bunker2023}, additional work is needed to calibrate metallicity diagnostics suitable for the study of galaxies at $z>10$.  Advances in this direction will be crucial for a reliable characterization of galaxies at the dawn of time over the next few years.

\section{Summary and Conclusions}
By using an analytical model of primordial dust emission,
we predict the {\em dust continuum FIR signatures} associated with massive galaxy sources similar to those recently detected by the JWST at $z\ga 7$.
Encouragingly, ALMA upper flux limits are consistent with our default model, which adopts currently expected properties for typical first galaxies (dust-to metal mass ratio: $D/M = 5 \times 10^{-3}$, gas metallicity: $Z_{\rm g} = 5 \times 10^{-3}~Z_\sun$, star formation efficiency: $\eta = 0.01$).
However, our model rules out very high $D/M \ga 0.06$ or high $Z_{\rm g} \ga 5\times 10^{-2}~Z_{\sun}$, in agreement with some scenarios proposed in the literature to explain the non-detection of the dust continuum for select JWST sources at $z \ga 10$.

According to our results, the determination of upper flux limits for the dust continuum emission can provide important clues to constrain the amount of dust and metal abundance in primeval galaxies (see Fig.~\ref{fig:fig1}). In addition, we also demonstrate that multi-band studies with ALMA would help to constrain the detailed chemistry and grain sizes of dust in the early universe (see Fig.~\ref{fig:fig2}). Overall, such constraints on the dust content of the first galaxies reflect the efficiency and characteristics of early metal enrichment, driven by the first generations of SN explosions \citep[e.g.,][]{karlsson2013,behrens2018}.

The powerful synergy between ALMA and JWST promises to provide a rich probe into the physical nature of the first galaxies, including the origin of dust during the initial stages of galaxy formation. A key challenge for future ALMA observations will be to reach lower flux limits, but also to increase the FIR wavelength coverage. This multi-waveband frontier may be extended to high-energy observations, as well, such as gamma-ray bursts triggered by the death of massive stars inside high-$z$ galaxies \citep[e.g.,][]{wang2012}. We clearly are entering a period of rapid discovery, promising to elucidate galaxies and their environment at the dawn of the universe.

\begin{acknowledgments}
We thank the referee for constructive comments and suggestions that improved this manuscript.
We also thank Alexander Ji for providing tabulated dust opacities for the different
dust models used here.
This work makes use of the Yggdrasil code \citep{zackrisson2011}, which adopts
Starburst99 SSP models, based on Padova-AGB tracks \citep{leitherer1999, vazquez2005}
for Pop~II stars.
\end{acknowledgments}

%

\vspace{5mm}





\bibliography{references}

\begin{thebibliography}{}
\expandafter\ifx\csname natexlab\endcsname\relax\def\natexlab#1{#1}\fi
\providecommand{\url}[1]{\href{#1}{#1}}
\providecommand{\dodoi}[1]{doi:~\href{http://doi.org/#1}{\nolinkurl{#1}}}
\providecommand{\doeprint}[1]{\href{http://ascl.net/#1}{\nolinkurl{http://ascl.net/#1}}}
\providecommand{\doarXiv}[1]{\href{https://arxiv.org/abs/#1}{\nolinkurl{https://arxiv.org/abs/#1}}}


\bibitem[{{Adams} {et~al.}(2023){Adams}, {Conselice}, {Ferreira}, {Austin},
  {Trussler}, {Juod{\v{z}}balis}, {Wilkins}, {Caruana}, {Dayal}, {Verma}, \&
  {Vijayan}}]{Adams2023}
{Adams}, N.~J., {Conselice}, C.~J., {Ferreira}, L., {et~al.} 2023, \mnras, 518,
  4755

\bibitem[{{Atek} {et~al.}(2023){Atek}, {Shuntov}, {Furtak}, {Richard}, {Kneib},
  {Mahler}, {Zitrin}, {McCracken}, {Charlot}, {Chevallard}, \&
  {Chemerynska}}]{Atek2022}
{Atek}, H., {Shuntov}, M., {Furtak}, L.~J., {et~al.} 2023, \mnras, 519, 1201

\bibitem[{{Bakx} {et~al.}(2022){Bakx}, {Zavala}, {Mitsuhashi}, {Treu},
  {Fontana}, {Tadaki}, {Casey}, {Castellano}, {Glazebrook}, {Hagimoto},
  {Ikeda}, {Jones}, {Leethochawalit}, {Mason}, {Morishita}, {Nanayakkara},
  {Pentericci}, {Roberts-Borsani}, {Santini}, {Serjeant}, {Tamura}, {Trenti},
  \& {Vanzella}}]{Bakx2022}
{Bakx}, T. J.~L.~C., {Zavala}, J.~A., {Mitsuhashi}, I., {et~al.} 2022, \mnras

\bibitem[{{Behrens} {et~al.}(2018){Behrens}, {Pallottini}, {Ferrara},
  {Gallerani}, \& {Vallini}}]{behrens2018}
{Behrens}, C., {Pallottini}, A., {Ferrara}, A., {Gallerani}, S., \& {Vallini},
  L. 2018, \mnras, 477, 552

\bibitem[{{Bianchi} \& {Schneider}(2007)}]{bianchi2007}
{Bianchi}, S., \& {Schneider}, R. 2007, \mnras, 378, 973

\bibitem[{{Boylan-Kolchin}(2022)}]{Boylan2022}
{Boylan-Kolchin}, M. 2022, arXiv:2208.01611

\bibitem[{{Bromm} \& {Yoshida}(2011)}]{BrommYoshida2011}
{Bromm}, V., \& {Yoshida}, N. 2011, \araa, 49, 373

\bibitem[{{Bunker} {et~al.}(2023){Bunker}, {Saxena}, {Cameron}, {Willott},
  {Curtis-Lake}, {Jakobsen}, {Carniani}, {Smit}, {Maiolino}, {Witstok},
  {Curti}, {D'Eugenio}, {Jones}, {Ferruit}, {Arribas}, {Charlot}, {Chevallard},
  {Giardino}, {de Graaff}, {Looser}, {Luetzgendorf}, {Maseda}, {Rawle}, {Rix},
  {Rodriguez Del Pino}, {Alberts}, {Egami}, {Eisenstein}, {Endsley},
  {Hainline}, {Hausen}, {Johnson}, {Rieke}, {Rieke}, {Robertson}, {Shivaei},
  {Stark}, {Sun}, {Tacchella}, {Tang}, {Williams}, {Willmer}, {Baker}, {Baum},
  {Bhatawdekar}, {Bowler}, {Boyett}, {Chen}, {Circosta}, {Helton}, {Ji}, {Lyu},
  {Nelson}, {Parlanti}, {Perna}, {Sandles}, {Scholtz}, {Suess}, {Topping},
  {Uebler}, {Wallace}, \& {Whitler}}]{bunker2023}
{Bunker}, A.~J., {Saxena}, A., {Cameron}, A.~J., {et~al.} 2023, arXiv e-prints,
  arXiv:2302.07256

\bibitem[{{Burkert}(1995)}]{burkert1995}
{Burkert}, A. 1995, \apjl, 447, L25

\bibitem[{{Cherchneff} \& {Dwek}(2010)}]{cherchneff2010}
{Cherchneff}, I., \& {Dwek}, E. 2010, \apj, 713, 1

\bibitem[{{Curtis-Lake} {et~al.}(2022){Curtis-Lake}, {Carniani}, {Cameron},
  {Charlot}, {Jakobsen}, {Maiolino}, {Bunker}, {Witstok}, {Smit}, {Chevallard},
  {Willott}, {Ferruit}, {Arribas}, {Bonaventura}, {Curti}, {D'Eugenio},
  {Franx}, {Giardino}, {Looser}, {L{\"u}tzgendorf}, {Maseda}, {Rawle}, {Rix},
  {Rodriguez del Pino}, {{\"U}bler}, {Sirianni}, {Dressler}, {Egami},
  {Eisenstein}, {Endsley}, {Hainline}, {Hausen}, {Johnson}, {Rieke},
  {Robertson}, {Shivaei}, {Stark}, {Tacchella}, {Williams}, {Willmer},
  {Bhatawdekar}, {Bowler}, {Boyett}, {Chen}, {de Graaff}, {Helton}, {Hviding},
  {Jones}, {Kumari}, {Lyu}, {Nelson}, {Perna}, {Sandles}, {Saxena}, {Suess},
  {Sun}, {Topping}, {Wallace}, \& {Whitler}}]{curtislake2023}
{Curtis-Lake}, E., {Carniani}, S., {Cameron}, A., {et~al.} 2022, arXiv
  e-prints, arXiv:2212.04568

\bibitem[{{De Rossi} \& {Bromm}(2017)}]{derossi2017}
{De Rossi}, M.~E., \& {Bromm}, V. 2017, \mnras, 465, 3668

\bibitem[{{De Rossi} \& {Bromm}(2019)}]{derossi2019}
---. 2019, \apj, 883, 113

\bibitem[{{De Rossi} {et~al.}(2018){De Rossi}, {Rieke}, {Shivaei}, {Bromm}, \&
  {Lyu}}]{derossi2018}
{De Rossi}, M.~E., {Rieke}, G.~H., {Shivaei}, I., {Bromm}, V., \& {Lyu}, J.
  2018, \apj, 869, 4

\bibitem[{{Ferrara} {et~al.}(2022){Ferrara}, {Pallottini}, \&
  {Dayal}}]{Ferrara2022}
{Ferrara}, A., {Pallottini}, A., \& {Dayal}, P. 2022, arXiv:2208.00720

\bibitem[{{Finkelstein} {et~al.}(2022){Finkelstein}, {Bagley}, {Arrabal Haro},
  {Dickinson}, {Ferguson}, {Kartaltepe}, {Papovich}, {Burgarella}, {Kocevski},
  {Huertas-Company}, {Iyer}, {Koekemoer}, {Larson}, {P{\'e}rez-Gonz{\'a}lez},
  {Rose}, {Tacchella}, {Wilkins}, {Chworowsky}, {Medrano}, {Morales},
  {Somerville}, {Yung}, {Fontana}, {Giavalisco}, {Grazian}, {Grogin}, {Kewley},
  {Kirkpatrick}, {Kurczynski}, {Lotz}, {Pentericci}, {Pirzkal}, {Ravindranath},
  {Ryan}, {Trump}, {Yang}, {Almaini}, {Amor{\'\i}n}, {Annunziatella},
  {Backhaus}, {Barro}, {Behroozi}, {Bell}, {Bhatawdekar}, {Bisigello}, {Bromm},
  {Buat}, {Buitrago}, {Calabr{\`o}}, {Casey}, {Castellano}, {Ch{\'a}vez Ortiz},
  {Ciesla}, {Cleri}, {Cohen}, {Cole}, {Cooke}, {Cooper}, {Cooray}, {Costantin},
  {Cox}, {Croton}, {Daddi}, {Dav{\'e}}, {de La Vega}, {Dekel}, {Elbaz},
  {Estrada-Carpenter}, {Faber}, {Fern{\'a}ndez}, {Finkelstein}, {Freundlich},
  {Fujimoto}, {Garc{\'\i}a-Argum{\'a}nez}, {Gardner}, {Gawiser},
  {G{\'o}mez-Guijarro}, {Guo}, {Hamblin}, {Hamilton}, {Hathi}, {Holwerda},
  {Hirschmann}, {Hutchison}, {Jaskot}, {Jha}, {Jogee}, {Juneau}, {Jung},
  {Kassin}, {Le Bail}, {Leung}, {Lucas}, {Magnelli}, {Mantha}, {Matharu},
  {McGrath}, {McIntosh}, {Merlin}, {Mobasher}, {Newman}, {Nicholls}, {Pandya},
  {Rafelski}, {Ronayne}, {Santini}, {Seill{\'e}}, {Shah}, {Shen}, {Simons},
  {Snyder}, {Stanway}, {Straughn}, {Teplitz}, {Vanderhoof}, {Vega-Ferrero},
  {Wang}, {Weiner}, {Willmer}, {Wuyts}, {Zavala}, \& {CEERS
  Team}}]{Finkelstein2022}
{Finkelstein}, S.~L., {Bagley}, M.~B., {Arrabal Haro}, P., {et~al.} 2022,
  \apjl, 940, L55

\bibitem[{{Fujimoto} {et~al.}(2022){Fujimoto}, {Finkelstein}, {Burgarella},
  {Carilli}, {Buat}, {Casey}, {Ciesla}, {Tacchella}, {Zavala}, {Brammer},
  {Fudamoto}, {Ouchi}, {Valentino}, {Cooper}, {Dickinson}, {Franco},
  {Giavalisco}, {Hutchison}, {Kartaltepe}, {Koekemoer}, {Kojima}, {Larson},
  {Murphy}, {Papovich}, {P{\'e}rez-Gonz{\'a}lez}, {Somerville}, {Yoon},
  {Wilkins}, {Yung}, {Akins}, {Amor{\'\i}n}, {Arrabal Haro}, {Bagley},
  {Chworowsky}, {Cooper}, {Costantin}, {Daddi}, {Ferguson}, {Grogin},
  {Jim{\'e}nez-Andrade}, {Juneau}, {Kirkpatrick}, {Kocevski}, {Le Bail},
  {Long}, {Lucas}, {Magnelli}, {McKinney}, {Rose}, {Seill{\'e}}, {Simons}, \&
  {Weiner}}]{Fujimoto2022}
{Fujimoto}, S., {Finkelstein}, S.~L., {Burgarella}, D., {et~al.} 2022,
  arXiv:2211.03896

\bibitem[{{Gall} {et~al.}(2011){Gall}, {Hjorth}, \& {Andersen}}]{Gall2011}
{Gall}, C., {Hjorth}, J., \& {Andersen}, A.~C. 2011, \aapr, 19, 43

\bibitem[{{Greif} \& {Bromm}(2006)}]{greif2006}
{Greif}, T.~H., \& {Bromm}, V. 2006, \mnras, 373, 128

\bibitem[{{Harikane} {et~al.}(2022{\natexlab{a}}){Harikane}, {Ouchi}, {Oguri},
  {Ono}, {Nakajima}, {Isobe}, {Umeda}, {Mawatari}, \& {Zhang}}]{Harikane2022}
{Harikane}, Y., {Ouchi}, M., {Oguri}, M., {et~al.} 2022{\natexlab{a}},
  arXiv:2208.01612

\bibitem[{{Harikane} {et~al.}(2022{\natexlab{b}}){Harikane}, {Inoue},
  {Mawatari}, {Hashimoto}, {Yamanaka}, {Fudamoto}, {Matsuo}, {Tamura}, {Dayal},
  {Yung}, {Hutter}, {Pacucci}, {Sugahara}, \& {Koekemoer}}]{Harikane2022b}
{Harikane}, Y., {Inoue}, A.~K., {Mawatari}, K., {et~al.} 2022{\natexlab{b}},
  \apj, 929, 1

\bibitem[{{Jaacks} {et~al.}(2018){Jaacks}, {Finkelstein}, \&
  {Bromm}}]{Jaacks2018}
{Jaacks}, J., {Finkelstein}, S.~L., \& {Bromm}, V. 2018, \mnras, 475, 3883

\bibitem[{{Ji} {et~al.}(2014){Ji}, {Frebel}, \& {Bromm}}]{ji2014}
{Ji}, A.~P., {Frebel}, A., \& {Bromm}, V. 2014, \apj, 782, 95

\bibitem[{{Kaasinen} {et~al.}(2022){Kaasinen}, {van Marrewijk}, {Popping},
  {Ginolfi}, {Di Mascolo}, {Mroczkowski}, {Concas}, {Di Cesare}, {Killi}, \&
  {Langan}}]{Kaasinen2022}
{Kaasinen}, M., {van Marrewijk}, J., {Popping}, G., {et~al.} 2022, arXiv
  e-prints, arXiv:2210.03754

\bibitem[{{Karlsson} {et~al.}(2013){Karlsson}, {Bromm}, \&
  {Bland-Hawthorn}}]{karlsson2013}
{Karlsson}, T., {Bromm}, V., \& {Bland-Hawthorn}, J. 2013, Reviews of Modern
  Physics, 85, 809

\bibitem[{{Klypin} {et~al.}(2021){Klypin}, {Poulin}, {Prada}, {Primack},
  {Kamionkowski}, {Avila-Reese}, {Rodriguez-Puebla}, {Behroozi}, {Hellinger},
  \& {Smith}}]{Klypin2021}
{Klypin}, A., {Poulin}, V., {Prada}, F., {et~al.} 2021, \mnras, 504, 769

\bibitem[{{Kohandel} {et~al.}(2023){Kohandel}, {Ferrara}, {Pallottini},
  {Vallini}, {Sommovigo}, \& {Ziparo}}]{Kohandel2023}
{Kohandel}, M., {Ferrara}, A., {Pallottini}, A., {et~al.} 2023, \mnras, 520,
  L16

\bibitem[{{Labbe} {et~al.}(2022){Labbe}, {van Dokkum}, {Nelson}, {Bezanson},
  {Suess}, {Leja}, {Brammer}, {Whitaker}, {Mathews}, {Stefanon}, \&
  {Wang}}]{Labbe2022}
{Labbe}, I., {van Dokkum}, P., {Nelson}, E., {et~al.} 2022, arXiv:2207.12446

\bibitem[{{Leitherer} {et~al.}(1999){Leitherer}, {Schaerer}, {Goldader},
  {Delgado}, {Robert}, {Kune}, {de Mello}, {Devost}, \&
  {Heckman}}]{leitherer1999}
{Leitherer}, C., {Schaerer}, D., {Goldader}, J.~D., {et~al.} 1999, \apjs, 123,
  3

\bibitem[{{Liu} \& {Bromm}(2022)}]{LiuBromm2022}
{Liu}, B., \& {Bromm}, V. 2022, \apjl, 937, L30

\bibitem[{{Mitchell-Wynne} {et~al.}(2015){Mitchell-Wynne}, {Cooray}, {Gong},
  {Ashby}, {Dolch}, {Ferguson}, {Finkelstein}, {Grogin}, {Kocevski},
  {Koekemoer}, {Primack}, \& {Smidt}}]{mitchell2015}
{Mitchell-Wynne}, K., {Cooray}, A., {Gong}, Y., {et~al.} 2015, Nature
  Communications, 6, 7945

\bibitem[{{Naidu} {et~al.}(2022){Naidu}, {Oesch}, {Setton}, {Matthee},
  {Conroy}, {Johnson}, {Weaver}, {Bouwens}, {Brammer}, {Dayal}, {Illingworth},
  {Barrufet}, {Belli}, {Bezanson}, {Bose}, {Heintz}, {Leja}, {Leonova},
  {Marques-Chaves}, {Stefanon}, {Toft}, {van der Wel}, {van Dokkum}, {Weibel},
  \& {Whitaker}}]{Naidu2022}
{Naidu}, R.~P., {Oesch}, P.~A., {Setton}, D.~J., {et~al.} 2022,
  arXiv:2208.02794

\bibitem[{{P{\'e}roux} \& {Howk}(2020)}]{peroux2020}
{P{\'e}roux}, C., \& {Howk}, J.~C. 2020, \araa, 58, 363

\bibitem[{{Planck Collaboration} {et~al.}(2014){Planck Collaboration}, {Ade},
  {Aghanim}, {Armitage-Caplan}, {Arnaud}, {Ashdown}, {Atrio-Barandela},
  {Aumont}, {Baccigalupi}, {Banday}, \& et~al.}]{planck2014}
{Planck Collaboration}, {Ade}, P.~A.~R., {Aghanim}, N., {et~al.} 2014, \aap,
  571, A16

\bibitem[{{Pollack} {et~al.}(1994){Pollack}, {Hollenbach}, {Beckwith},
  {Simonelli}, {Roush}, \& {Fong}}]{pollack1994}
{Pollack}, J.~B., {Hollenbach}, D., {Beckwith}, S., {et~al.} 1994, \apj, 421,
  615

\bibitem[{{Popping}(2023)}]{Popping2023}
{Popping}, G. 2023, \aap, 669, L8

\bibitem[{{Popping} \& {P{\'e}roux}(2022)}]{popping2022}
{Popping}, G., \& {P{\'e}roux}, C. 2022, \mnras, 513, 1531

\bibitem[{{Smith} {et~al.}(2019){Smith}, {Ma}, {Bromm}, {Finkelstein},
  {Hopkins}, {Faucher-Gigu{\`e}re}, \& {Kere{\v{s}}}}]{Smith2019}
{Smith}, A., {Ma}, X., {Bromm}, V., {et~al.} 2019, \mnras, 484, 39

\bibitem[{{V{\'a}zquez} \& {Leitherer}(2005)}]{vazquez2005}
{V{\'a}zquez}, G.~A., \& {Leitherer}, C. 2005, \apj, 621, 695

\bibitem[{{Wang} {et~al.}(2012){Wang}, {Bromm}, {Greif}, {Stacy}, {Dai},
  {Loeb}, \& {Cheng}}]{wang2012}
{Wang}, F.~Y., {Bromm}, V., {Greif}, T.~H., {et~al.} 2012, \apj, 760, 27

\bibitem[{{Yoon} {et~al.}(2022){Yoon}, {Carilli}, {Fujimoto}, {Castellano},
  {Merlin}, {Santini}, {Yun}, {Murphy}, {Jung}, {Casey}, {Finkelstein},
  {Papovich}, {Fontana}, {Treu}, \& {Letai}}]{Yoon2022}
{Yoon}, I., {Carilli}, C.~L., {Fujimoto}, S., {et~al.} 2022, arXiv:2210.08413

\bibitem[{{Zackrisson} {et~al.}(2011){Zackrisson}, {Rydberg}, {Schaerer},
  {{\"O}stlin}, \& {Tuli}}]{zackrisson2011}
{Zackrisson}, E., {Rydberg}, C.-E., {Schaerer}, D., {{\"O}stlin}, G., \&
  {Tuli}, M. 2011, \apj, 740, 13

\end{thebibliography}
\bibliographystyle{aasjournal}



\end{document}